\documentclass[cite,epsf,12pt]{article}
\usepackage{graphicx}

\newcommand{\beq}{\begin{equation}}
\newcommand{\eeq}{\end{equation}}
\newcommand{\beqarray}{\begin{eqnarray}}
\newcommand{\eeqarray}{\end{eqnarray}}

\begin{document}


\hyphenation{Ginz-burg Le-van-yuk  com-pound multilayered 
ap-pli-ca-tion con-sid-er com-pound su-per-con-duc-ting}

\newcommand{\ie}{{i.e.}}
\newcommand{\eg}{{e.g.}}
\newcommand{\etal}{{\it et al.}}
\newcommand{\eq}[1]{~(\ref{#1})}
\newcommand{\eqdos}[2]{~(\ref{#1},\ref{#2})}
\renewcommand{\epsilon}{\varepsilon}
\newcommand{\lsim}{\stackrel{<}{_\sim}}
\newcommand{\gsim}{\stackrel{>}{_\sim}}

\newcommand{\Tc}{\mbox{$T_c$}}
\newcommand{\Tco}{\mbox{$T_{c0}$}}
\newcommand{\TcH}{\mbox{$T_{c}(H)$}}
\newcommand{\Ds}{\mbox{$\Delta\sigma$}}
\newcommand{\Dchi}{\mbox{$\Delta\chi$}}
\newcommand{\DM}{\mbox{$\Delta M$}}
\newcommand{\Hcdoso}{\mbox{$H_{c2}(0)$}}
\newcommand{\hc}{\mbox{$h^C$}}
\newcommand{\hC}{\hc}
\renewcommand{\epsilon}{\varepsilon}


\newcommand{\titulo}{
Breakdown by a magnetic field\\ of the superconducting fluctuations\\ in the 
normal state:\\
A simple phenomenological explanation
}

\newcommand{\autor}{M.V. Ramallo, J. Mosqueira, C. Carballeira, F. Soto, F. Vidal\footnote{Corresponding author ({\tt fmvidal@usc.es}; fax +34 981531682; tel.~+34 981563100 ext.14031).}}

\newcommand{\direccion}{
Laboratorio de Baixas Temperaturas e
Superconductividade\\  (Unidad Asociada al ICMM-CSIC,
Spain),\\  Departamento de F\'{\i}sica da Materia
Condensada,\\  Universidade de Santiago de Compostela,
E-15782, Spain}

\hrule  
\begin{center}
\mbox{}\vspace{-0.5cm}\\
{\nonfrenchspacing\it 
As~published~in~J.~Phys.~Chem.~Solids~{\bf 67},~479-481~(2006)\\
\small(Proceedings of SNS'04, Sitges, Spain)}
 \end{center}
  \hrule \mbox{}\\
\mbox{}\vspace{-1cm}\\ 
\begin{center}
  \Large\bf
\titulo\\  \end{center}\mbox{}\vspace{-1cm}\\

\begin{center}\normalsize\autor\end{center} 

\begin{center}\normalsize\it\direccion\end{center}


\mbox{}\vskip0.5cm{\bf Abstract. }
We first summarize our recent observations, through magnetization 
measurements in different low-$T_c$ superconductors, of a rather sharp 
disappearance of the superconducting fluctuations (SCF) in the normal state when 
the magnetic field approaches \Hcdoso, the upper critical field extrapolated 
to $T=0$~K. We propose that a crude phenomenological description of the observed effects may be obtained if the quantum limits associated with the uncertainty principle are introduced in the Gaussian-Ginzburg-Landau description of the fluctuation-induced magnetization.

\newpage
\setlength{\baselineskip}{18pt}


\section{Introduction}

One of the better ``spectroscopic'' methods to obtain  various central 
characteristic parameters of superconductors (including their superconducting coherence length)
is the study of the superconducting thermal fluctuations (SCF) above  the
superconducting transition. The Cooper pairs created by SCF produce contributions
in the  normal state to, among other properties, the magnetization
(resulting  in the so-called fluctuation-induced magnetization \DM).  In both low- and high-\Tc\ superconductors, those contributions can be measured and then analyzed in terms of various 
models, notably the Gaussian-Ginzburg-Landau (GGL) approach. Various reviews of this type of experiments may be seen, \eg, in 
\cite{ST,VidalReview}. However, an interesting aspect only recently  addressed (in
particular, experimentally\cite{PRBRC}) is the behaviour of the SCF under
strong magnetic fields [for reduced-fields $h\equiv H/\Hcdoso$ well above 0.2, 
being \Hcdoso\ the Ginzburg-Landau amplitude of the upper magnetic field
extrapolated to
$T=0$K]. Specifically, it was addressed the question of up to what magnetic fields
do the SCF exist. For that, the fluctuation-induced magnetization was measured under strong
magnetic fields  in various superconducting Pb-In alloys.\cite{PRBRC}  The first aim of the present Communication is to
summarize some of these recent experimental results, with emphasis on the findings that there is a reduced-magnetic field
$\hc\simeq1.1$ above which the SCF are no longer observed, and that \DM\ decreases with $h$
very rapidly upon approaching that \hc-value (faster than any previous theory
of the SCF under magnetic fields). 

The second aim of this Communication is
to propose a crude phenomenological explanation for such a high-$h$ behaviour of the SCF. This will
be based on the introduction in the GGL approach of the idea that the uncertainty principle imposes a limit to the shrinkage of the superconducting wave function when the magnetic field increases. This will extend to high fields our previous proposals for the SCF at high reduced-temperatures,\cite{VidalEPL} in spite that the magnetic field is an antisymmetric perturbation\cite{ST}. In fact, our results suggest the existence of an unexpected ``quantum protectorate''\cite{QP} for the coherent fluctuating Cooper pairs above $H_{c2}(T)$, that is only broken by the limits imposed by the uncertainty principle.

\section{Some experimental results in Pb-In alloys under high magnetic fields}

Some examples of our measurements of \DM\  in the normal state in Pb-In alloys are summarized in figs.~1(a) and (b).  The details of these measurements may be seen in \cite{PRBRC}. Figure 1(a) shows that $\Delta M(h)_\varepsilon$ depends linearly on $h$   for $h\ll\epsilon$, where $\epsilon\equiv\ln(T/\Tco)$ is the reduced-temperature and \Tco\ the zero-field critical temperature.  This is the behaviour predicted for such a low-$h$ regime by the pioneering calculations of Schmidt and Schmid\cite{SS}. When $h\gsim0.2$, as may be seen in figs.~1(a) and (b),  $\Delta M(h)_\varepsilon$ begins to decrease and for $h\gsim\hC\simeq1.1$ the fluctuation-induced diamagnetism vanishes. The sharp high-$h$ decay of the SCF is especially evident in Fig.~1(b). There it may be seen also  that the general form and location of this  decay is essentially independent on the dirtiness of the superconductor.  In particular,  the $\DM(h)_{T_{c0}}$  curves when normalized as suggested by the Prange predictions\cite{Prange} (see also below) are the same for all the studied superconductors, except for pure Pb where \mbox{$\DM(h)_{T_{c0}}$} also  decays strongly upon approaching $\hC\simeq1.1$ in spite of its different low-$h$ amplitude. We note that the low-$h$  \DM\ differences  between the pure Pb and the Pb-In alloys are today well known to be due to nonlocal effects, which in clean superconductors may reduce \DM\ somewhat even in the Schmidt limit $h\ll\epsilon$, while they do not affect \DM\ significantly in unclean superconductors.\cite{ST}

\begin{figure}[h]
\mbox{}\hfill
\includegraphics[scale=0.44]{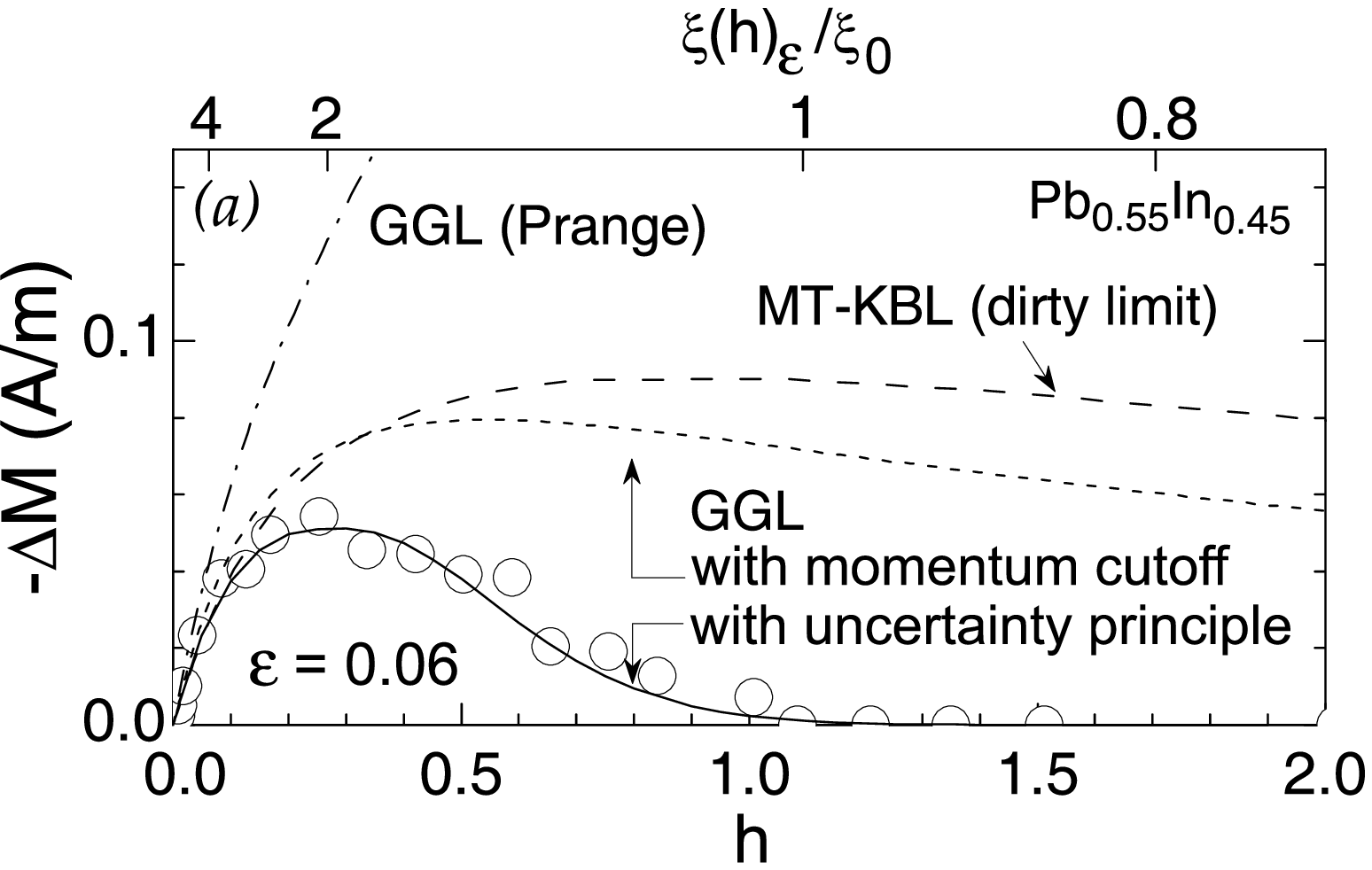}\mbox{}\hspace{12pt}\mbox{}
\includegraphics[scale=0.55]{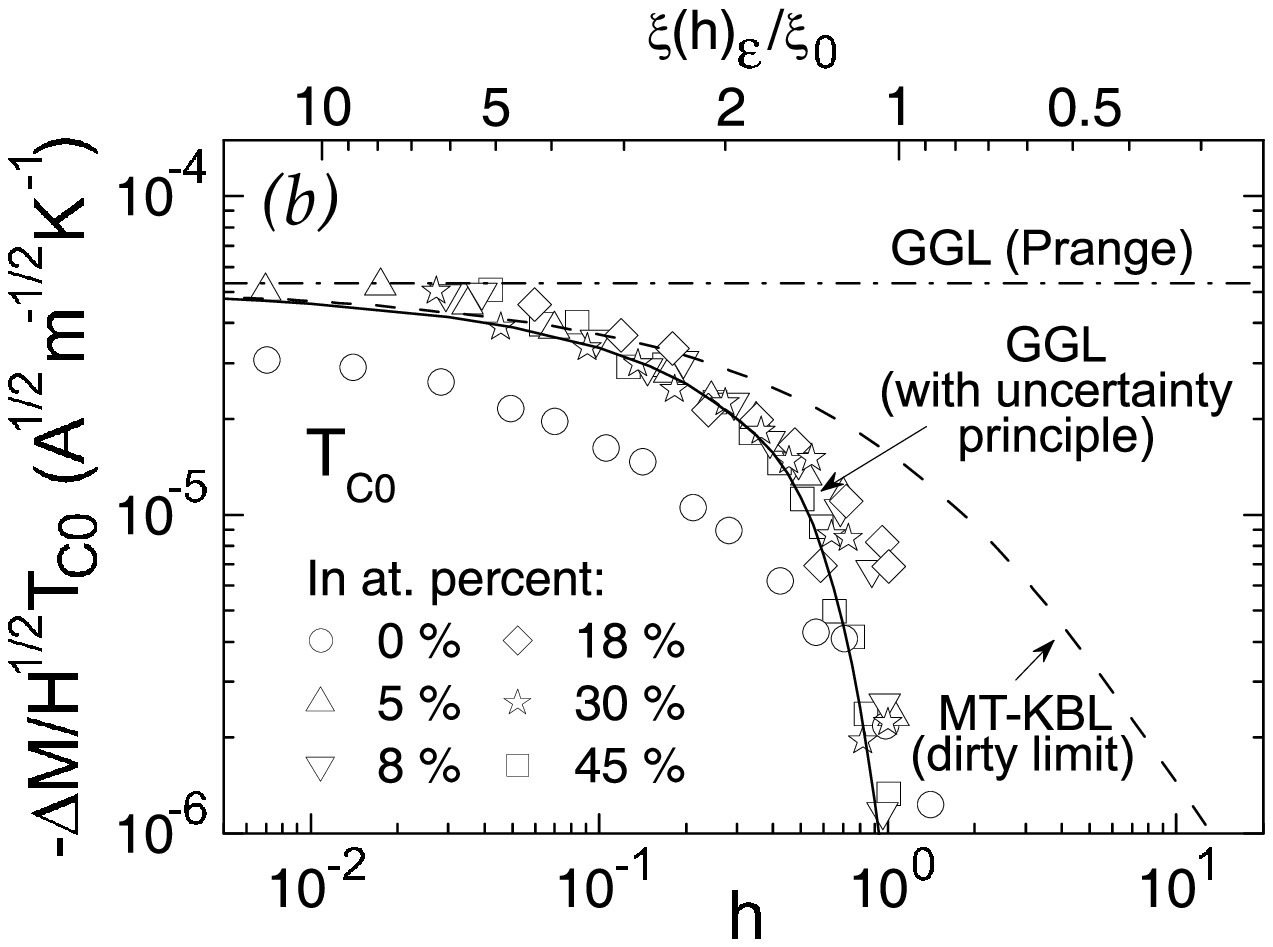}\vspace{-0.6cm}\mbox{}
\hfill\mbox{}\\
\caption{Fluctuation-induced magnetization versus reduced-magnetic field,  {\it (a)} in the 0.45 at.\%~In alloy at a reduced-temperature $\epsilon=0.06$, and {\it (b)} in all the measured Pb-In alloys at the zero-field critical temperature and  normalized to the Prange value\cite{Prange}. See main text and Ref.~\cite{PRBRC} for details.\vspace{12pt}\mbox{}}
\end{figure}

\section{Failure of some previous calculations of  \DM\ to explain the $h\gsim0.2$  data }
In figs.~1(a) and (b) we have plotted the predictions resulting from the main previous calculations of \DM\ under  finite magnetic fields.\cite{Prange,MT,KBL,Momento} The dot-dashed lines correspond to the early Prange calculation,\cite{Prange} consisting on the application of the Schmid's GGL formalism\cite{SS} to  finite fields. These calculations do not include the short-wavelength effects induced by the magnetic field.\cite{ST,MT,KBL} In the case of dirty superconductors, an introduction of those effects was done by Maki and Takayama\cite{MT} and by Klemm, Beasley and Luther (MT-KBL)\cite{KBL} on the grounds of the BCS microscopic theory, by considering the dynamics of the SCF.  In figs.~1(a) and (b), we have plotted as dashed lines these MT-KBL predictions. Also plotted, as dotted lines, are the GGL calculations obtained in \cite{Momento} by introducing in the Prange formalism a momentum (or kinetic-energy) cutoff of the fluctuation modes. As visible in these figures, none of these calculations agree, not even qualitatively, with the observed decay of \DM\ for $h\gsim0.2$. They also do not predict the existence of a maximum $h$ value for the occurrence of SCF. As it could be expected, the Prange predictions\cite{Prange} are the ones with a lower $h$-range of validity. However, we note that the MT-KBL predictions agree well  with the  data of the unclean materials up to $h\simeq0.2$, including their $\DM/H^{1/2}\Tco$ independence of In concentration.

\section{A simple phenomenological GGL explanation of the breakdown of \DM\  at high $h$.}
A remarkable aspect of the experimental results summarized above, that may provide a hint for extensions of the calculations of \DM\ to the high-field regime, is that, as illustrated in figs.~1(a) and (b), for the field amplitudes where \mbox{$\Delta M(h)_\varepsilon$} vanishes the GL coherence length, $\xi(h)_\epsilon$, becomes of the order of $\xi_0$, the actual superconducting coherence length at $T=0$~K. The $\xi(h)_\varepsilon/\xi_0$ scale in figs.~1(a) and (b) was obtained by using $\xi(h)_\varepsilon=\sqrt{2}\xi(0)h^{-1/2}$\cite{ST} and also $\xi(0)=0.74\xi_0$, which is still a good approximation in the dirty limit.\cite{ST} When compared with our previous results at low fields but at high reduced-temperatures\cite{VidalEPL}, this last finding already suggests that, in spite of the antisymmetric character of the magnetic field, the vanishing of $\Delta M(h)$ may also be due to the quantum constraints to the shrinkage of the superconducting wave function:\cite{VidalEPL} Even above \Tco, the superconducting coherence length cannot be smaller than its minimum value, given by the uncertainty principle, the one at $T=0$~K (which, in fact, is the characteristic length of the Cooper pairs\cite{ST}). When  the shrinkage of the superconducting wave function is due to a magnetic field, this condition may be written as $\xi(h)_\varepsilon\stackrel{>}{_\sim}\xi_0$,
where $\xi_0$ for each alloy is related to the one of pure Pb, $\xi_0^{Pb}$, by\cite{ST} $\xi_0\simeq(\xi_0^{Pb}\ell)^{1/2}$, $\ell$ being the electronic mean free path. Such inequality directly leads to a critical reduced-field, $h^C$, given by $h^C=2(\xi(0)/\xi_0)^2$, above which all the SCF vanish. By using again $\xi(0)=0.74\xi_0$, we obtain $h^C\simeq1.1$, in excellent agreement with the results of figs.~1(a) and (b). As $\xi(0)/\xi_0$ is almost material-independent,\cite{ST} the relationship $\xi(h)_\varepsilon\stackrel{>}{_\sim}\xi_0$ predicts that the above value of $h^C$ will be ``universal'', in  strikingly good agreement with the experimental results at \Tco\ for all the samples studied in this work and summarized in Fig.~1(b).

To qualitatively estimate below \hC\ the effects on \DM\ of the uncertainty-principle limitations, probably the simplest way is to introduce them in the conventional GGL framework, in spite that the latter formally applies only near the transition (we stress that extensions of the GGL approach beyond its formal application range have been already done, using different approximations, by many workers\cite{ST}). For that, we first note that in terms of the ``total energy'' $E_{nk_z}$ of the fluctuating modes of Landau level index $n=0,1...$ and wave vector parallel to the field $k_z$, this constraint may be written as:
\begin{equation}
E_{nk_z}\equiv\epsilon+(2n+1)h+\xi^2(0)k_z^2\;\lsim\;(\xi(0)/\xi_0)^2,
\label{cutH}
\end{equation}
where the energies are expressed in units of  $\hbar^2/2m^*\xi^2(0)$, and $\hbar$ and $m^*$ are, respectively, the Planck constant and the effective mass of the Cooper pairs. 
Note that this inequality was already introduced in previous GGL calculations of \DM\cite{Energia} for low magnetic fields by cutting off the statistical sums over the Landau-level index. The Landau index $n$ was there considered as a continuous variable and Eq.~(1) was taken as an all-or-nothing prohibition for the existence of the fluctuating modes.\cite{Energia} This so-called total-energy cutoff procedure is appropriate for low reduced-fields ($h\lsim0.2$), but it becomes inadequate as $h$ increases and Eq.~(1) is fulfilled by a smaller number of Landau levels: For instance, for $(\xi(0)/\xi_0)^2=0.55$, $\epsilon=0.03$ and $h=0.25$ the higher $n$-value allowed by Eq.~(1) would be $n\simeq0.5$, which must be rounded $\sim50$\%  
to be consistent with the discreteness of $n$. Note that, in fact, the final \DM\ expressions proposed in \cite{Energia} do not predict any vanishing of the fluctuation magnetization for magnetic fields above $\hC\simeq1.1$. To solve such difficulties, we introduce a total-energy-dependent weighting function, $W(E_{nk_z})$, pondering the contribution of each fluctuating mode in the free-energy statistical sum. This procedure is similar to the one proposed by Patton, Ambegaokar and Wilkins (PAW).\cite{PAW} However, PAW's approach does not take into account the limits imposed by the uncertainty principle to the shrinkage of the superconducting wave function. In fact, PAW's calculations are equivalent to the choice $W_{\rm PAW}(E_{nk_z})=\ln[1+\exp(-E_{nkz}/h_0)]/\ln E_{nk_z}$, which does not consider the inequality (1). Here the reduced-field $h_0$ corresponds to the maximum of the $\Delta M(h)_{T_{C0}}$ curve, and therefore in our Pb-In alloys it will be $0.2\lsim h_0\lsim0.25$. In our present study, to reproduce the rapid fall-off of the SCF expected when the inequality (1) begins to be violated, we introduce an additional prefactor to the penalization function, using $W(E_{nk_z})=W_{\rm PAW}(E_{nk_z})(1+\exp[(E_{nk_z}-(\xi(0)/\xi_0)^2-\delta)/\delta])^{-1}$. This additional prefactor has the form of a Fermi-Dirac distribution function, having a step-like decay starting at energies $\sim(\xi(0)/\xi_0)^2$ and with $\delta$ as typical half-width. By repeating the standard GGL calculations for $\DM(T,H)$ in isotropic 3D superconductors above the transition\cite{Prange} but now including the weighting function $W(E_{nk_z})$, we obtain:
\begin{equation}
\DM=
{\frac{k_{\rm B} T} {\pi\phi_0}}
\int_{0}^{\infty}{\rm d}k_z\sum_{n=0}^\infty{\frac{\partial}{\partial h}}\left[
W(E_{nk_z}) h \ln E_{nk_z}\right],
\label{Mfl}
\end{equation}
where $k_B$ is the Boltzmann constant and $\phi_0$ the flux quantum. This formula may be numerically computed thanks to the rapid decay of $W(E_{nk_z})$ as $n$ or $k_z$ increase. In figs.~1(a) and (b) we plot that evaluation as a solid line, using again  $(\xi(0)/\xi_0)^2=0.55$. We also used $h_0=0.22$ and $\delta=0.2$,  the values giving a better agreement with experiments. As may be seen in the figures, this agreement is excellent in the unclean superconductors for all $h$, including also  the vanishing of $\Delta M(h)_\varepsilon$ at $h^C\simeq1.1$.  
In the case of pure Pb, although our expressions qualitatively reproduce the main features of its high-$h$ \DM, it fails to account for the \DM\ amplitude al low fields, due to the not consideration of the nonlocal effects affecting this superconductor. 
Finally, we emphasize again that in spite of its success our present crude GGL approach  must be seen only as an argument indicating the relevance of quantum effects in the high-field SCF of Pb-In alloys, and not at all as a {\it real theory} of such effects.

\section{Conclusions}
Our measurements of \DM\ in Pb-In alloys up to magnetic fields above \Hcdoso\ show a rapid decrease of the SCF effects for reduced-fields $h\gsim0.2$, and their vanishing for $h$ of the order of  1.1. 
A crude phenomenological description of these effects was obtained by introducing in the Gaussian-Ginzburg-Landau description of \DM\  the limits associated with the uncertainty principle to the shrinkage of the superconducting wave function.
Our results suggest then that \DM\ at high reduced-fields is dominated by the uncertainty principle constraints, the antisymmetric character of the magnetic field playing a much less relevant role. An interesting question which remains open is the possible relationship between the vanishing of the SCF when $\xi(h)_\epsilon\simeq\xi_0$ and the MT-KBL microscopic approach\cite{MT,KBL} if zero-point contributions are considered.

\mbox{}

\mbox{}

\mbox{}\\ {\Large \bf Acknowledgements}\\ \mbox{}\\
\nonfrenchspacing
We acknowledge support from the ESF ``Vortex'' Program, the CICTY, Spain, under Grant MAT 2004-04364,  and Uni\'on Fenosa under Grant 220/0085-2002.

\newpage


\begin{thebibliography}{99}
%
\nonfrenchspacing \setlength{\parskip}{0pt}

\newcommand{\revista}[4]{\nonfrenchspacing{\nonfrenchspacing\it #1}\/ {\bf #2}, #3 (#4)}
\renewcommand{\autor}[2]{\nonfrenchspacing#1 #2}
\newcommand{\y}{and }
\newcommand{\PRB}{Phys.~Rev.~B}


\bibitem{ST} See, \eg, \autor{M.}{Tinkham}, {\it Introduction to Superconductivity} (McGraw-Hill, New York, 1996), chaps.~4 and~8.

\bibitem{VidalReview}
\autor{F.}{Vidal}  and \autor{M.V.}{Ramallo}, in {\it The gap symmetry and fluctuations in high-\Tc\ superconductors} (NATO-ASI series, Plenum, 1998), ed. J. Bok, G. Deutscher, D. Pavuna, and S.A. Wolf.


\bibitem{PRBRC} F. Soto \etal,  \revista{\PRB}{70}{060501(R)}{2004}.



\bibitem{VidalEPL} \autor{F.}{Vidal} \etal, \revista{Europhys. Lett.}{59}{754}{2002}, and references therein.

\bibitem{QP} \autor{R.B.}{Laughlin} \y \autor{D.}{Pines}, \revista{Proc.~Natl.~Acad.~Sci.~U.S.A.}{97}{28}{2000}; \autor{P.W.}{Anderson}, \revista{Science}{288}{480}{2000}.


\bibitem{SS}
\autor{H.}{Schmidt}, \revista{Z. Phys.}{216}{336}{1968};
\autor{A.}{Schmid}, \revista{Phys. Rev.}{180}{527}{1969}.

\bibitem{Prange} \autor{R.E.}{Prange}, \revista{Phys. Rev. B}{1}{2349}{1970}.

\bibitem{MT} \autor{K.}{Maki} \y \autor{H.}{Takayama}, \revista{J. Low Temp. Phys.}{5}{313}{1971}; 
\autor{K.}{Maki}, \revista{Phys. Rev. Lett.}{30}{648}{1973}.

 \bibitem{KBL}\autor{R.A.}{Klemm}, \autor{M.R.}{Beasley}, \y \autor{A.}{Luther}, \revista{\PRB}{8}{5072}{1973}.

\bibitem{Momento}\autor{C.}{Carballeira} \etal,
\revista{Phys.$\,$Rev.$\,$Lett.$\mbox{}$}{84}{3157}{2000}; \revista{Physica C}{384}{185}{2003}.


\bibitem{Energia}
\autor{J.}{Mosqueira}, \autor{C.}{Carballeira}, \autor{F.}{Vidal}, \revista{Phys. Rev. Lett.}{87}{167009}{2001}.

\bibitem{PAW} \autor{B.R.}{Patton}, \autor{V.}{Ambegaokar}, \y \autor{J.W.}{Wilkins}, \revista{Solid State Comm.}{7}{1287}{1969}; \autor{B.R.}{Patton} \y \autor{J.W.}{Wilkins}, \revista{\PRB}{6}{4349}{1972}.

\end{thebibliography}
\end{document}